\documentclass[doublecol]{epl2} 
\usepackage{subfigure}
\usepackage{color}
\def\be{\begin{equation}}
\def\ee{\end{equation}}

\def\bc{\begin{center}} 
\def\ec{\end{center}}
\def\bea{\begin{eqnarray}}
\def\eea{\end{eqnarray}}
\newcommand{\avg}[1]{\langle{#1}\rangle}

\title{Connect and win: The role of social networks in political elections}
\shorttitle{Connect and win: The role of social networks in political elections}

\author{Arda Halu\inst{1}, Kun Zhao\inst{1}, Andrea Baronchelli\inst{1, 2}, Ginestra Bianconi\inst{3}}
\shortauthor{A. Halu \etal}

\institute{                    
  \inst{1}   Department of Physics, Northeastern University, Boston, 
Massachusetts 02115 USA.\\
\inst{2} Laboratory for the Modeling of Biological and Socio-technical Systems,\\ Northeastern University, Boston MA 02115 USA\\
\inst{3} School of Mathematical Sciences, Queen Mary University of London, London E1 4NS, United Kingdom}


\abstract{Many real systems are made of strongly interacting networks, with profound consequences on their dynamics. Here, we consider the case of two interacting social networks and, in the context of a simple  model, we address the case of political elections. Each network represent{s} a competing party and every agent, on the election day, can choose to be either active in one of the two networks (vote for the corresponding party) or to be inactive in both (not vote). The  opinion dynamics during the election campaign is described through a simulated annealing algorithm. We find that for a large region of the parameter space the result of the competition between the two parties allows for the existence of pluralism in the society, where both parties have a finite share of the votes. The central result is that a densely connected social network is key for the final victory of a party. However, small committed minorities can play a crucial role, and even reverse the election outcome.}

\begin{document}

\maketitle

\section{Introduction}
Interacting and interdependent networks have recently {attracted} great attention \cite{Havlin1,Vespignani,Havlin2,Havlin3,Grassberger,Dorogovtsev,Yamir1, Yamir2,Jesus,Ivanov}. Here, the function of a node in one network depends on the operational level of the { nodes it is dependent on} in the other networks. Investigated examples {range} from infrastructure networks as the power-grid and the Internet \cite{Havlin1} to interacting biological networks in physiology \cite{Ivanov}. Understanding how critical phenomena \cite{crit, Dynamics} are affected by the presence of interaction{s} or interdependent networks is crucial to control and monitor the dynamics of and on complex systems.
In this context it was shown that interdependent networks are more fragile than single networks, and that the percolation transitions can be first order \cite{Havlin1}.

Interesting, but so far less explored, is the case of interacting social networks, describing individuals that manage their personal relationships in different social contexts (e.g., work, family, friendship, etc.). Taking into account these multiple layers is crucial, as proven recently for community detection methods in social networks \cite{Fortunato_com, Lambiotte, Lehmann}, but the effect of their presence is still not understood in many respects. For example, there is considerable current interest in opinion models \cite{Fortunato}, among which we cite the the Sznajd model \cite{Sznajd}, the voter model \cite{Redner},  the naming game \cite{Andrea,Korniss} and Galam models \cite{Galam,Galam_05}. {But the influence of more than one network has gathered less attention\cite{galam_percolation}.}


Here we propose a simple model for opinion dynamics that describes two parties competing for votes during a political campaign. Every opinion, i.e., party, is modeled as a social network through which a contagion dynamics can take  place. Individuals, on the other hand, are represented by a node on each network, and can be active only in one of the two networks (vote for one party) at the moment of the election. Each agent has also a third option \cite{Andrea,Lama,ng_opinion,Korniss,Strogatz}, namely not to vote, and in that case she will be inactive in both networks. Crucially, agents are affected by the opinion of their neighbors, and the nodes tend to be active in the networks where their neighbors are also active. Moreover, the chance of changing opinion decreases as the decision moment approaches, in line with the observation that vote preferences stabilize as the election day comes closer \cite{polls}.

Our aim is to provide insights in the role of multiple social networks in the voting problem through a simple and clear mathematical model, in the spirit, for example, of recent work concerning the issue of ideological conflict \cite{Strogatz}. We  describe the dynamics of social influence in the two networks, and we model the uncertainty reduction preceding the vote through a simulated annealing process. {Long} before the election the agents change opinions and can sustain a small fraction of antagonistic relations, but as the election approaches their dynamics slows down, until they reach the state in which the dynamics is frozen, at the election day. At that moment, the party winning the elections is the one with more active nodes. Finally, we focus on the case in which the networks sustaining each party are represented by two Poisson graphs, and address the role of different average connectivities. This choice is consistent for example with the data on social networks of mobile phone communication, {which} are characterized by a typical scale in the degree (being fitted with a power-law distribution of  exponent $\gamma=8.4$) \cite{Onnela}.

We observe a rich phase diagram  of the opinion dynamics.
The results are that in the thermodynamic limit the most connected network wins the election {independent of} the initial condition of the system, in agreement with recent results on the persuasive role of a densely connected social network \cite{centola}. However, for a large region of the parameters the voting results of the two parties are very close and small perturbations could alter the results. In this context, we observe that a small minority of committed agents can reverse the outcome of the election result, thus confirming the results obtained in very recent and different models \cite{Korniss,Strogatz}.

\section{Parties as antagonistic social networks}
We consider two antagonistic networks $A, B$ representing the social networks of two competing political parties.
Each agent $i$ is represented in each network and can choose to be active in one of the networks.
In particular   $\sigma^A_i=0$ if agent $i$ is inactive in network $A$ and $\sigma_i^A=1$ if agent $i$ is active in network $A$.
Similarly $\sigma_i^B=0,1$ indicates if a node is active or {inactive} in network $B$.
Since ultimately the activity of an individual in a network corresponds to the agent voting {for} the corresponding party, each agent can be active only on one network on the election day (i.e. if $\sigma_i^A=1$ then $\sigma_i^B=0$ and if $\sigma_i^B=1$ then $\sigma_i^A=0$).
Nevertheless we leave to the agent the freedom not to vote, in that case $\sigma_i^A=\sigma_i^B=0$.
 Moreover agents are influenced by  their neighbors. 
 Therefore, we assume that,  {on} the election day, if at least one neighbor of agent $i$ is active  in network $A$, the agent will be active in the same network (network A) \textit{provided that it is not already active in network $B$}.
We assume that a  symmetrical  process is  occurring for the opinion dynamics in network B.
Hence,  the mathematical constraints that our agent opinions need to satisfy at the election day are:
\begin{eqnarray}
\sigma_i^A&=&\left[1-\prod_{j\in N_A(i)} (1-\sigma_j^A)\right](1-\sigma_i^B)\nonumber \\
\sigma_i^B&=&\left[1-\prod_{j\in N_B(i)} (1-\sigma_j^B)\right](1-\sigma_i^A),
\label{cons}
\end{eqnarray}
where $N_A(i)$ ($N_B(i)$) are the set of neighbors of node $i$ in network A (network B).
Therefore at the election day people cannot anymore change their opinion. On the contrary before the election we allow for some conflicts in the system, and in general the constraints provided by Eqs.~$\ref{cons}$ will not be satisfied.
\begin{figure}
\begin{center}
\includegraphics[width=0.6\columnwidth]{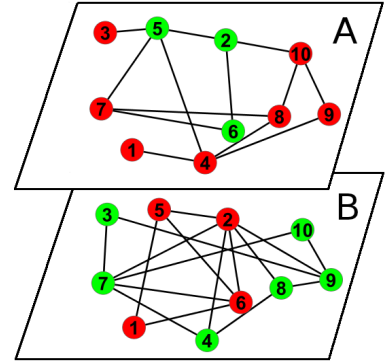}
\end{center}
\caption{(Color online) The two competing political parties are represented by two networks. Each agent is  represented in both networks but can {either be active (green node) in only one of the two or inactive (red node) in both networks.} Moreover the activity of neighbor nodes influences the opinion of any given node.}
\label{fig1}
\end{figure}

\section{Evolution dynamics during the election campaign}

To model how agents decide {on} their vote during the pre-election period we consider the following  algorithm.
We consider a Hamiltonian that counts the number of the constraints in Eq. (\ref{cons}) that are violated.
Therefore we take a Hamiltonian $H$ {of} the following form
\begin{eqnarray}
H&=&\sum_i\left\{\sigma_i^A-\left[1-\prod_{j\in N_A(i)} (1-\sigma_j^A)\right](1-\sigma_i^B)\right\}^2+\nonumber \\
&&\sum_i\left\{\sigma_i^B-\left[1-\prod_{j\in N_B(i)} (1-\sigma_j^B)\right](1-\sigma_i^A)\right\}^2.
\end{eqnarray}
{The terms in the brackets can take on the values $\pm1, 0$, therefore a natural choice of Hamiltonian to count the number of constraint violations involves squares of these terms.}

\begin{figure}
\begin{center}
\includegraphics[width=0.8\columnwidth, height=11cm]{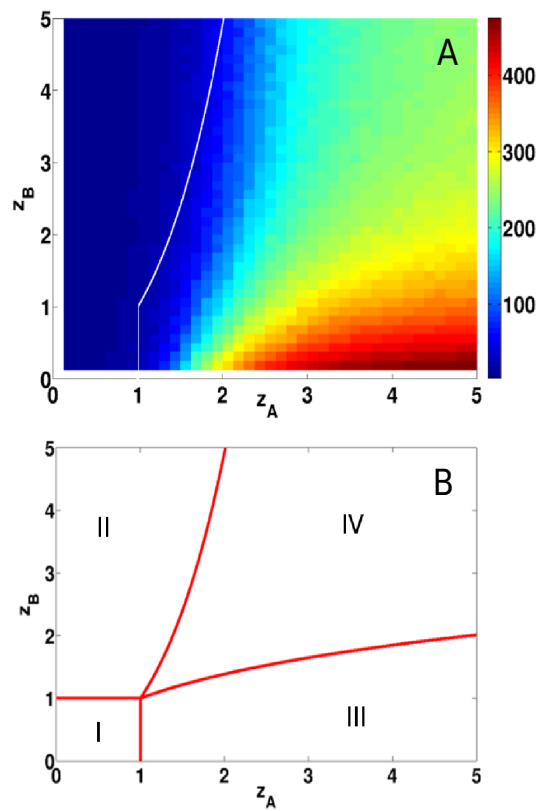}
\end{center}
\caption{(Color online) (Panel A) The size of the largest connected component $S_A$ in network A at the end of the simulated annealing calculation as a function of the average connectivity of the two networks: $z_A$ and $z_B$ respectively. The data is simulated for two networks for $N=500$ nodes and averaged 60 times. The simulated annealing algorithm is independent {of} initial conditions. The white line represent the boundary between the region in which network A is percolating and the region in which network A is not percolating. (Panel B) The schematic representation of the different phases of the proposed model. In region I none of the networks is percolating, in region II network B is percolating, in region III network  A is percolating, in region IV both networks are percolating. }
\label{SA}
\end{figure}

\begin{figure}
\begin{center}
\includegraphics[width=\columnwidth]{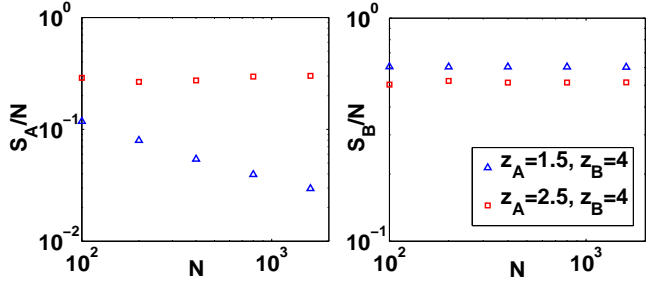} 
\end{center}
\caption{(Color online) We represent the fraction of nodes in the giant component $S_A$ of network A and in the giant component $S_B$ of network B in different regions of the phase space.
In region II ($z_A=1.5, z_B=4$) the giant component in network A ($S_A$ ) disappear{s} in the thermodynamic limit  while in region IV ($z_A=2.5, z_B=4$) it remains constant. The giant component in network B remains constant in the thermodynamic limit both in region II and in region IV.
 Each data point  is simulated for the two networks for $N$ nodes and averaged 200 times.}
\label{finite}
\end{figure}

\begin{figure}
\begin{center}
\includegraphics[width=0.8\columnwidth, height=6.0cm ]{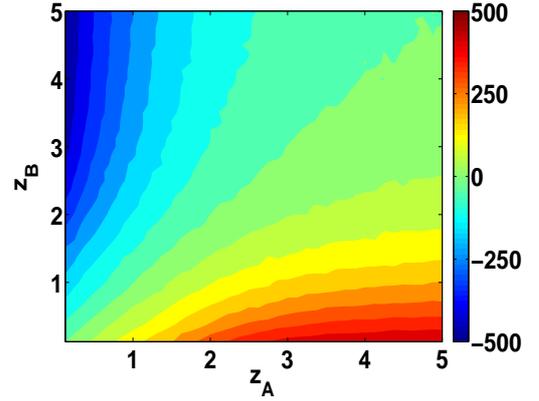}
\end{center}
\caption{(Color online)  The contour plot for the difference between the total number of votes $m_A$ in party A (total number of agents active in network A) {and} the total number of votes $m_B$ in party B (total number of agents active in network B).   The data is simulated for two networks for $N=500$ nodes and averaged 90 times. {It is clear that the larger the difference in average connectivity of the two networks, the larger the advantage of the more connected political party}. }
\label{majority}
\end{figure}

We start from {an} initial condition {where the active nodes in networks A and B are distributed uniformly randomly}, and we consider the fact that {long} before the election the agents are free to change opinion. Therefore we model their dynamics as a Monte Carlo dynamics which  equilibrate{s} following the  Hamiltonian $H$  {starting from} a relative{ly} high initial temperature, i.e. {initially} some conflicts are allowed in the system. {Therefore, initially  the active nodes in networks A and B are distributed according to the high temperature Gibbs measure, mimicking an effectively ``unbiased'' population at the beginning of  the campaigning process. Moreover we note here that since we start with a sufficiently high temperature, the dynamics is not affected by the specific initial conditions of the system.} 
As the election day approach{es},  the effective temperature of the opinion dynamics decreases and the agents tend to reduce to zero the number of conflicts with their neighbors.
The opinion dynamics described in this way is implemented with a simulated annealing algorithm.
We start at { a} temperature $T=1$ and we allow the system to equilibrate by $2N$  Monte Carlo steps where a node is picked randomly in either {one} of the networks with equal probability and {is changed from active to inactive or vice versa. Subsequently, the Hamiltonian, or the number of conflicts, is recalculated. If the opinion flip results in a smaller number of conflicts, it is accepted. Else, it is accepted with probability $e^{-\Delta H /kT}$}. {This Monte Carlo process is repeated by slowly reducing} the temperature by a multiplicative factor {of} $0.95$ until we reach the  temperature state $T=0.01$  where the Hamiltonian is $H=0$, there are no more conflicts in the network, and the probability of one spin flip is about $e^{-1/0.01}\simeq 10^{-44}$. {The choice of increment in the temperature reduction is such that the overall simulation time is compatible with the dynamics of social systems. The Monte Carlo sweeps that are performed, each of which corresponds to one campaigning day, span a total number of $\log 0.01 / \log 0.95 \approx 90$ days.}
It turns out that the Hamiltonian $H$ has in general multiple fundamental states and the simulated annealing algorithm always find one of these states.
The final configuration for the model just described  is depicted in Figure $\ref{fig1}$.
In Figure $\ref{SA}$ we report the result of this opinion dynamics for two antagonistic networks {A, B} with Poisson degree distribution{s} and different average connectivities {$z_A$, $z_B $, respectively}.
In particular we plot the size $S_A$ of the giant component of the percolating cluster in network A, {i.e. the largest connected component of active nodes in network A}.
Additionally we have  characterized the  finite size effects (see Figure $\ref{finite}$) and concluded  that the phase diagram of the model is consistent with the following scenario valid in the  limit of large network sizes:
\begin{itemize}
\item
{\it Region (I):
$z_A<1, z_B<1$
}.
In this region both giant components in network A ($S_A$) and network B ($S_B$) are zero, $S_A=0, S_B=0$, and therefore essentially agents never vote.
\item
{\it Region (II) in Figure $\ref{SA}$:
}
In this region the giant component in network {B} emerges, $S_B>0, S_A=0$.
\item
{\it Region (III) in Figure $\ref{SA}$:
}
In this region the giant component in network {A} emerges, $S_A>0, S_B=0$.
\item
{\it Region (IV) in Figure $\ref{SA}$}
In this region we have the pluralism solution of the opinion dynamics and both giant component in networks A and B are different from zero, $S_A>0, S_B>0$.
\end{itemize}
In Regions II (III) the active agents in party B (party A) percolate the system while agents in party A (party B) remain concentrated in disconnected clusters.
Nevertheless, if the average connectivity of the two antagonistic parties is comparable (Region IV), the system can sustain an effective pluralism of opinions with both parties percolating in the system.
Therefore, we find the interesting result that if the connectivity of the two parties is large enough,i.e. we are in region IV of the phase diagram (Figure $\ref{SA}B$) the pluralism can be preserved in the model and there will be two parties with {a } high number of votes.
In order for a party to win the election, it is necessary that the active agents percolate in the corresponding network. The election outcome, nevertheless, depends crucially on the total number of votes in network A, $m_A$ and the total number of votes in network B, $m_B$.
In Figure $\ref{majority}$ we plot the difference between the number of votes in network A and the number of votes in network B. Very interestingly, we observe that the more connected party (network) has the majority of the votes. It is also worth noting that the final outcome of the election does not depend on the initial conditions. Overall, this result support{s} the intuition that if a party has a supporting network that is more connected it will win the elections, and is coherent with recent results concerning the role of densely connected social networks on the adoption of a behavior \cite{centola}.

 
\section{Committed agents}
\begin{figure}
\begin{center}

\includegraphics[width=\columnwidth]{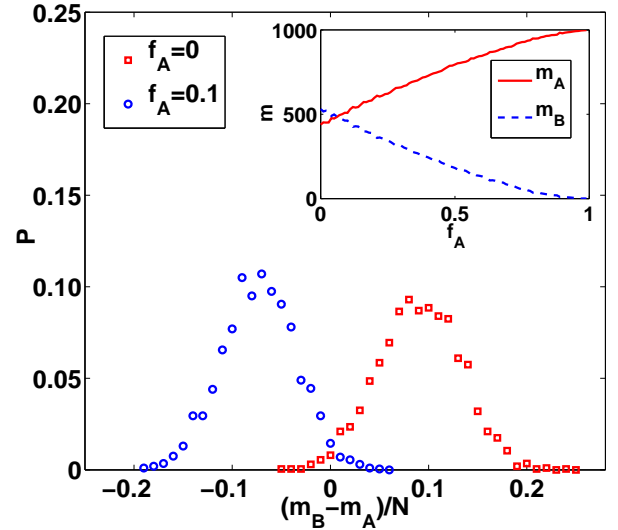} 
\end{center}
\caption{(Color online) We represent  the role of a  fraction $f$ of committed agents in reverting the outcome of the election. In particular we plot the histogram of the difference between the  {fraction of agents $m_B/N$ voting for party $B$ and the fraction of agents $m_A/N$ voting for party $A$} for a fraction $f_A$ of committed agents to party A, with $f_A=0$ and $f_A=0.1$ and average connectivities  of the networks $z_A=2.5, z_B=4$. The histogram is performed for 1000 realizations of two networks of size $N=1000$.  In the inset we  show the average {number} of agents in network A ($m_A$) and agents in network B ($m_B$) as a function of the fraction of committed agents $f_A$. A small  fraction of agents ($f_A\simeq 0.1$) is sufficient to reverse the outcome of the elections. The data in the inset is simulated for two networks for $N=1000$ nodes and averaged 10 times.}
\label{committed}
\end{figure}

Different opinion-dynamics models have recently considered the role of committed agents \cite{Strogatz,Korniss,galam_committed}. Here we explore the effect of committed individuals during the election campaign by considering a situation in which a fraction of the nodes {always remain} active in one of the two networks, never {changing their} opinion.
Figure \ref{committed} shows that in Region IV a small fraction of agents $f\simeq 0.1 $ \textit{in the less connected network} can reverse the outcome of the election. Indeed the probability distribution $P=P(m_A-m_B)$ in different realization of the dynamics is shifted towards the party supported by the committed minority. Remarkably, this finding fits perfectly with the results of the radically different models proposed in \cite{Strogatz,Korniss}, and generalizes them to the case of political elections. The crucial role potentially played by committed minorities is thus suggested by different models in different aspects of social dynamics, suggesting the need for future work exploring these findings.



\section{Conclusions}
In conclusion, we have put forth a simple model for the opinion dynamics taking place during an election campaign. We have modeled parties (or opinions) in terms of a social networks, and individuals in terms of nodes belonging to these social networks and connecting them. We have considered the case of antagonistic agents who have to decide for a single party, or for none of them. We have described the quenching of the opinions preceding the voting moment as a simulated annealing process where the temperature is progressively lowered till the voting moment, when the individuals minimize the number of conflicts with their neighbors. We have shown that there is a wide region in the phase diagram where two antagonistic parties survive gathering a finite fraction of the votes, and therefore {the existence of} pluralism in the election system. Moreover, we have pointed out that a key quantity to get a finite share of the overall votes is the connectivity of the networks corresponding to  different parties. Nevertheless connectivity is not sufficient to win the elections, since a small fraction of committed agents is sufficient to invert the results of the voting {process}. 

Though deliberately basic, the model provides insights into different aspects of the election dynamics. {Moreover, from a broader perspective, our work proposes a general framework for the description of any opinion formation process involving different contexts/networks, where opinions are frozen at some point in time, and where the agents' behavior reflects the approach of that point such that they are initially less susceptible to influence from their neighborhoods (high initial temperatures) and attempt to reduce the level of frustration/conflict more strongly later (low temperatures).} In {future works} we plan to generalize the model by studying antagonistic networks with different topolog{ies}, such as competing scale-free and Poisson networks or two competing scale-free networks. Other extensions of this mode{l} could describe several competing parties, consider a threshold dynamics as the one triggering the opinion formation of the agents in \cite{centola}, or relax the hypothesis of purely antagonistic interactions, thus allowing the agents to express multiple preferences in a multi-layered opinion space.


\end{document}